\documentclass[aps,showpacs,superscriptaddress,preprint]{revtex4}%
\usepackage{amsfonts}
\usepackage{amsmath}
\usepackage{amssymb}
\usepackage{graphicx}%
\providecommand{\U}[1]{\protect\rule{.1in}{.1in}}
\providecommand{\U}[1]{\protect\rule{.1in}{.1in}}
\providecommand{\U}[1]{\protect\rule{.1in}{.1in}}
\providecommand{\U}[1]{\protect\rule{.1in}{.1in}}
\providecommand{\U}[1]{\protect\rule{.1in}{.1in}}
\providecommand{\U}[1]{\protect\rule{.1in}{.1in}}
\providecommand{\U}[1]{\protect\rule{.1in}{.1in}}
\providecommand{\U}[1]{\protect\rule{.1in}{.1in}}

\begin{document}
\title{Atomic hydrogen adsorption and incipient hydrogenation of the Mg(0001)
surface: A density-functional theory study }
\author{Yanfang Li}
\affiliation{College of Materials Science and Engineering, Taiyuan University of
Technology, Taiyuan 030024, People's Republic of China}
\author{Ping Zhang}
\thanks{Corresponding author; zhang\_ping@iapcm.ac.cn}
\affiliation{LCP, Institute of Applied Physics and Computational Mathematics, P.O. Box
8009, Beijing 100088, People's Republic of China}
\author{Bo Sun}
\affiliation{LCP, Institute of Applied Physics and Computational Mathematics, P.O. Box
8009, Beijing 100088, People's Republic of China}
\author{Yu Yang}
\affiliation{LCP, Institute of Applied Physics and Computational
Mathematics, P.O. Box 8009, Beijing 100088, People's Republic of
China}
\author{Yinghui Wei }
\affiliation{College of Materials Science and Engineering, Taiyuan University of
Technology, Taiyuan 030024, People's Republic of China}


\begin{abstract}
We investigate the atomic hydrogen adsorption on Mg($0001$) by using
density-functional theory within the generalized gradient approximation and a
supercell approach. The coverage dependence of the adsorption structures and
energetics is systematically studied for a wide range of coverage $\Theta$
(from 0.11 to 2.0 monolayers) and adsorption sites. In the coverage range 0%
$<$%
$\Theta\mathtt{<}$1.0, the most stable among all possible adsorption sites is
the on-surface fcc site followed by the hcp site, and the binding energy
increases with the coverage, thus indicating the higher stability of
on-surface adsorption and a tendency to the formation of H islands (clusters)
when increasing the coverage within the region 0%
$<$%
$\Theta\mathtt{<}$1.0. The on-surface diffusion path energetics of atomic
hydrogen, as well as the activation barriers for hydrogen penetration from the
on-surface to the subsurface sites, are also presented at low coverage. At
high coverage of 1.0%
$<$%
$\Theta\mathtt{\leq}$2.0, it is found that the coadsorption configuration with
1.0 monolayer of H's residing on the surface fcc sites and the remaining
($\Theta-$1.0) monolayer of H's occupying the subsurface tetra-I sites is most
energetically favourable. The resultant H-Mg-H sandwich structure for this
most stable coadsorption configuration displays similar spectral features to
the bulk hydride MgH$_{2}$ in the density of states. The other properties of
the H/Mg(0001) system, including the charge distribution, the lattice
relaxation, the work function, and the electronic density of states, are also
studied and discussed in detail. It is pointed out that the H-Mg chemical
bonding during surface hydrogenation displays a mixed ionic/covalent character.

\end{abstract}
\maketitle

\section{Introduction}

The interaction between the hydrogen and the magnesium has long received many
experimental
\cite{Hanada2004,Liang1999,Shang2004,Zaluska2005,Oelerich2001,Pelletier2001,Barkhordarian2004,Fatay2005,Pozzo2008,Renner1978,Topler1982,Spatz1993,Plummer1994,Vegge2005}%
\ and theoretical
\cite{Vegge2005,Jacobson2002,Noritake2002,Ohba2004,Wu2008,Song2004,Zhang2007,Staikov1999,Li2006,Wachowicz2001,Norskov1981,Johansson2006,Badescu2003}
concerns, especially due to the fact that hydrogen is the best candidate of
clean fuel in the future and Mg is an important potential hydrogen storage
material. One key blockade to prevent Mg from practical application in
hydrogen storage is its high reaction temperature and kinetic barrier for
hydrogenation and dehydrogenation \cite{Bobet2000}. However, recent systematic
experiments \cite{Hanada2004,Liang1999,Shang2004,Song2004,Vegge2005} have
shown that the hydrogen reaction kinetics can be prominently improved by
adding transition elements (such as Ni, Ti) as catalyst in the magnesium
hydride (MgH$_{2}$). Along this line many interesting phenomena have been
found, for example, it has been shown that alloying Mg with Ni can weaken the
bonding between H and Mg atoms and thus favor H$_{2}$ adsorption/dissociation process.

On the other hand, a thorough understanding of the atomic H adsorption on the
Mg surface from basic quantum-mechanical viewpoint is still lacking, although
some scarce theoretical \cite{Vegge2005, Jacobson2002, Noritake2002, Ohba2004,
Wu2008} and experimental \cite{Renner1978,Topler1982, Spatz1993, Plummer1994,
Vegge2005} data have been existed in previous reports. In particular, the
coverage dependence of the Mg-H bonding properties is highly interesting but
remains yet to be fully studied. Motivated by this observation, in this paper,
we present a first-principles study by systematically calculating the coverage
dependence of atomic hydrogen structures on the Mg(0001) surface and
subsurface in a variety of coverage range from 0.11 to 2.0 monolayer (ML).
Through these energetics calculations, we analyze and discuss the atomic
hydrogen adsorption structures, the Mg-H chemical bonding properties, and the
surface diffusion and penetration energetics of atomic hydrogen on Mg(0001).
By comparing the binding energies and the adsorption behavior at different
coverages, we obtain the saturation coverage of 1.0 ML on surface and find the
spontaneously relaxed H-Mg-H sandwich structure at the coverage larger than
1.0 ML. On the whole, our present systematic first-principles study provides a
detailed enlightening information of the incipient hydrogenation of Mg(0001).

The rest of this paper is organized as follows: In Sec. II, we describe our
first-principle calculation method and the models used in this paper. In Sec.
III, we present in detail our calculated results, including the on-surface and
subsurface H adsorption energies in a wide range of coverage, the structures
and properties of the H-Mg bonds, and the surface diffusion and penetration
energetics. At last, the conclusion is given in Sec. IV.

\section{Computational method}

The first-principles calculations based on the density functional theory (DFT)
are performed using the Vienna \textit{ab initio} simulation package
\cite{VASP} (VASP) with the projector-augmented-wave (PAW) pseudopotentials
\cite{PAW} and plane waves. The plane-wave energy cutoff was set to $250$ eV.
The so-called\textit{ repeated slab} geometries are applied. This scheme
consists of the construction of a unit cell of an arbitrarily fixed number of
atomic layers identical to that of the bulk in the plane of the surface
(defining the bidimensional cell), but symmetrically terminated by an
arbitrarily fixed number of empty layers (the \textquotedblleft\textit{vacuum}%
\textquotedblright) along the direction perpendicular to the surface. In the
present study, the clean Mg($0001$) surface is modeled by periodic slabs
consisting of nine magnesium layers separated by a vacuum of 20 \AA , which is
found to be sufficiently convergent \cite{Zhang2007}. The hydrogen atoms are
adsorbed on both sides of the slab in a symmetric way. During our
calculations, the outermost three magnesium layers, as well as the H atoms,
are allowed to relax while the central three layers of the slab are fixed in
their calculated bulk positions. If not mentioned differently we have used a
(15$\times$15$\times$1) $k$-point grid for the $p$($1\times1$) surface cell,
(9$\times$9$\times$1) $k$-point grid for the $p$($\sqrt{3}\times\sqrt{3}$) and
$p$($2\times2$) cells, and (7$\times$7$\times$1) $k$-point grid for the
$p$($3\times3$) cell, with Monkhorse-Pack scheme \cite{Pack}. Furthermore, the
generalized gradient approximation (GGA) of Perdew \textit{et al}.
\cite{GGA-2} for the exchange-correlation potential is employed. A Fermi
broadening \cite{Fermi broaden} of $0.02$ eV is chosen to smear the occupation
of the bands around $E_{F}$ by a finite-$T$ Fermi function and extrapolating
to $T$=$0$ K. \begin{figure}[ptb]
\includegraphics[angle=0,width=0.5\textwidth]{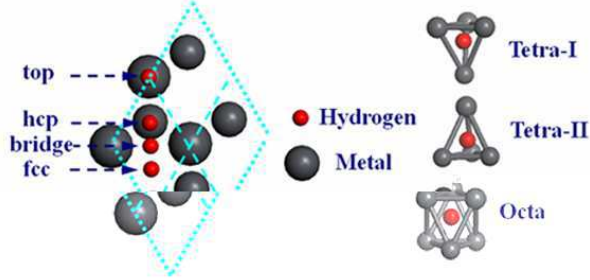} \caption{(Colour
online). (Left panel) Four on-surface adsorption sites including fcc, hcp,
bridge and on-top sites. (Right panel) Three subsurface adsorption sites
including tetra-I, tetra-II and octa sites. Note that Mg atoms of out layers
are shown by scaled grey balls.}%
\label{fig1}%
\end{figure}

In the present paper, the calculations for hydrogen atoms in the seven
adsorption sites, including on-surface (top, hcp, fcc, and bridge) and
subsurface (tetra-I, tetra-II, and octa) sites depicted in Fig. 1, have been
performed for coverage ranging from 0.11 to 2.0 ML. Specially, the hydrogen
coverage of 0.11 ML and 0.33 ML are calculated using $p$($3\times3$)\ surface
unit cell, while the coverage of 0.25, 0.5, 0.75, 1.0, 1.25, 1.5, 1.75 and 2.0
ML\ are calculated in the $p$($2\times2$) surface cell. The on-surface top and
bridge adsorption sites are found to be unstable by the fact that the hydrogen
atom initially located on these two sites will always move to the fcc site
after relaxation. The subsurface tetra-I site at low coverage is also found to
be unstable against the relaxation. Actually, the H atom placed on this
subsurface site will penetrate upward to a surface site after relaxation. Thus
in this paper, most of the on-surface adsorption studies are focused on the
fcc and hcp sites, while for the subsurface adsorption, the tetra-II and octa
sites are mainly considered.

One central quantity tailored for the present study is the average binding
energy of the adsorbed hydrogen atom defined as
\begin{equation}
E_{b}(\Theta)=-\frac{1}{N_{\text{H}}}[E_{\text{H/Mg($0001$)}}%
-E_{\text{Mg($0001$)}}-\text{ }N_{\text{H}}E_{\text{H}}],
\end{equation}
where $N_{\text{H}}$ is the total number of H adatom present in the
supercell at the considered coverage $\Theta$ (we define $\Theta$ as
the ratio of the number of adsorbed atoms to the number of atoms in
an ideal substrate layer). $E_{\text{H/Mg(0001)}}$,
$E_{\text{Mg(0001)}}$, and $E_{\text{H}}$ are the total energies of
the slabs containing hydrogen, of the corresponding clean Mg(0001)
slab, and of a free (spin polarized) hydrogen atom, respectively.
Thus a positive value of $E_{b}$ indicates that the adsorption is
exothermic (stable) with respect to a free H atom and a negative
value indicates endothermic (unstable) reaction. On the other hand,
since in most cases, the hydrogen chemisorption process inevitably
involves the dissociation of H$_{2}$ molecules, thus the adsorption
energy per hydrogen atom can alternatively be referenced to the
energy which the H atom has in the H$_{2}$ molecule by subtracting
half the dissociation energy $D$ of the H$_{2}$ molecule,
\begin{equation}
E_{\text{ad(}1/2\text{H}_{2}\text{)}}=E_{b}-D/2.
\end{equation}
With this choice of adsorption energy, then a positive value indicates that
the dissociative adsorption of H$_{2}$ is an exothermic process, while a
negative value indicates that it is endothermic and that it is energetically
more favorable for hydrogen to be in the gas phase as H$_{2}$.

\section{Results and discussion}

\subsection{Bulk Mg and clean Mg(0001) surface}

Before studying the H adsorption on the Mg($0001$) surface, we first consider
the bulk Mg and clean Mg(0001) surface. Our calculated lattice parameters for
bulk hcp Mg are $a$=3.207 \AA \ and $c/a$=$1.60$, well comparable to the
experimental values of 3.21 \AA \ and 1.62 \cite{Amonenko1962,Ashcroft1976}.
The calculated cohesive energy is $-$1.52 eV/atom, also in good agreement with
the experimental value of $-$1.51 eV/atom \cite{Amonenko1962,Ashcroft1976}.
The orbital-resolved electronic density of states (DOS) per atom for the bulk
Mg is shown in Fig. 2(a) \begin{figure}[ptb]
\includegraphics[angle=0,width=0.5\textwidth]{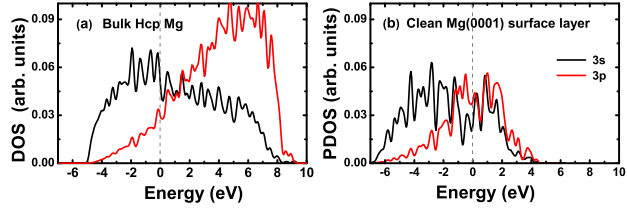}\caption{(Color
online). (a) The orbital-resolved DOS of bulk Mg and (b) the orbital-resolved
site-projected DOS for top layer of the clean p(1$\times$1)-Mg($0001$) film.
The Fermi energy is set at zero.}%
\label{fig2}%
\end{figure}with the Fermi energy set at zero. The two broad peaks correspond
to Mg $3s$ and $3p$ states, which are heavily mixed each other near the Fermi
energy. This $s$-$p$ orbital mixing results from the characteristic hcp
structure of bulk Mg in its ground state. In addition, the distribution form
of the total DOS in a wide energy range of $-$5.0 eV$\mathtt{<}E\mathtt{<}$4.0
eV can be nearly considered as a function of $E^{1/2}$, which is also a
typical feature of the $sp$-hybrid simple metals.

\begin{table}[th]
\caption{The calculated results of the surface energy $E_{s}$ (eV) and the
work function $\Phi$ (eV) for the clean Mg($0001$) surfaces in different
$k$-point meshes. }%
\label{Table1}
\begin{tabular}
[c]{cccc}\hline\hline
model & $\ k$-point mesh \  & $\ E_{s}$ (eV) \  & $\ \Phi$ (eV)\\\hline
1 $\times$ 1 & $12\times12\times1$ & 0.2911 & 3.711\\
1 $\times$ 1 & $15\times15\times1$ & 0.2947 & 3.729\\
2 $\times$ 2 & $7\times7\times1$ & 0.2892 & 3.717\\
2 $\times$ 2 & $9\times9\times1$ & 0.2977 & 3.733\\
$\sqrt{3}\times\sqrt{3}$ & $7\times7\times1$ & 0.2881 & 3.701\\
$\sqrt{3}\times\sqrt{3}$ & $9\times9\times1$ & 0.2951 & 3.728\\
3 $\times$ 3 & $7\times7\times1$ & 0.2936 & 3.723\\
3 $\times$ 3 & $9\times9\times1$ & 0.2911 & 3.704\\\hline\hline
\end{tabular}
\end{table}

The calculation for the atomic relaxations of the clean surface with
$1\mathtt{\times}1$, $\sqrt{3}\mathtt{\times}\sqrt{3}$, $2\mathtt{\times}2$,
and $3\mathtt{\times}3$ periodicities provides not only a test of the clean
surface with different cell sizes, but is also used to evaluate the charge
density difference used later and assess the changes in the work function by
hydrogen adsorption. Our clean-surface calculation shows that the two
outermost Mg($0001$) layers relax significantly from the bulk values. The
first-second interlayer expansion is nearly $2\%$ and the second-third
interlayer expansion is about $0.58\%$, which compares well with recent
first-principles results \cite{Wu2008,Wachowicz2001}. Note that the first
interlayer separation on most metal surfaces is contracted. From this aspect,
Mg($0001$) is an anomalous example showing the outermost interlayer expansion,
which has been proved and interpreted by Staikov \textit{et} \textit{al}.
\cite{Staikov1999}. The calculated charge density $n(\mathbf{r})$ (not
depicted here) of the clean Mg(0001) surface shows that similar to the other
typical metal surfaces \cite{Krak1981}, there is a rapid variation in
$n(\mathbf{r})$ in the surface interstitial region, with $n(\mathbf{r})$
falling off sharply in magnitude toward the vacuum and soon \textquotedblleft
healing\textquotedblright\ the discrete atomic nature. This sizable charge
redistribution near the surface is associated with the formation of the
(uniform) surface dipole layer, which sensitively determines the work
function. As one knows, the surface calculation requires enough $k$-point
meshes, efficient energy cutoff, the correct model, and the other details to
be the minimum numerical errors. To test the convergency of the physical
properties of the clean Mg($0001$) surface, we have calculated the surface
energy $E_{s}$ and the work function $\Phi$ of the clean Mg($0001$) slabs by
using different models with various $k$-point meshes. The results are listed
in Table I, from which it reveals that the influences to the surface
energetics from using different models are negligibly small. To keep the
computation accuracy as high as possible for reliable comparison between
different adsorption configurations, here we take to calculate and analyze
surface energetics and electronic structures under the same model. For
example, a $p(3\mathtt{\times}3)$ surface unit cell is needed if one wants to
analyze the properties of surface chemical activity\ in a coverage range
beginning\ from $\Theta$=0.11.

Figure 2(b) plots the orbital-resolved site-projected density of states (PDOS)
for the topmost Mg layer of the clean $p$($1\mathtt{\times}1$)\ Mg($0001$)
surface cell. Compared to Fig. 2(a), one can see that the surface electronic
structure of the clean Mg($0001$) differs from its bulk counterpart by a
downward shift in PDOS and a decreasing charge occupation of the Mg $3p$
states. Note that throughout this paper, we do not consider the quantum size
effects on the atomic and electronic structures, since in our various
supercell models the substrate has been fixed with the same thickness.

\begin{figure}[pth]
\includegraphics[angle=0,width=0.4\textwidth]{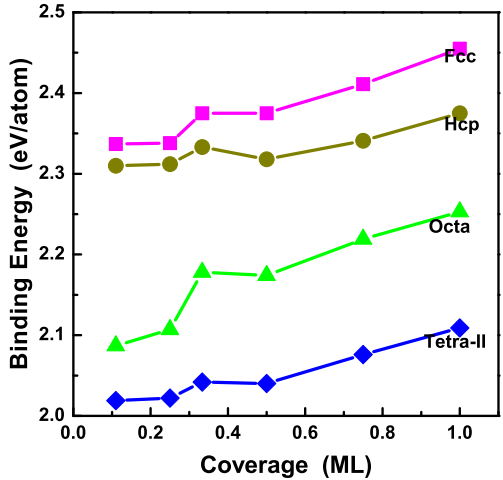}\caption{(Color
online). Calculated binding energy $E_{b}$ of H/Mg(0001) system versus the
coverage for the H atom adsorption in different sites. The solid lines
connecting the calculated binding energies are used to guide the eyes.}%
\end{figure}

\subsection{Atomic hydrogen adsorption at $\Theta\mathtt{\leq}$1.0}

\begin{table}[th]
\caption{ The calculated binding energy E$_{b}$ (in eV) and work
function $\Phi$ (in eV) as functions of atomic hydrogen coverage
on different sites of
Mg(0001) surface.}%
\begin{tabular}
[c]{cccccccc}\hline\hline & site & $\Theta$=0.11 & \ 0.25 \ \  & \
\ 0.33 \ \  & \ \ 0.50 \ \  & \ \ \ 0.75 \ \  & \ 1.0 \ \\\hline
& Fcc & 2.337 & 2.338 & 2.375 & 2.375 & 2.411 & 2.455\\
E$_{b}$ & Hcp & 2.320 & 2.324 & 2.324 & 2.326 & 2.332 & 2.375\\
& Octa & 2.020 & 2.022 & 2.035 & 2.050 & 2.086 & 2.109\\
& Tetra-II & 2.047 & 2.217 & 2.218 & 2.224 & 2.230 & 2.209\\\hline
& Fcc & 3.664 & 3.584 & 3.570 & 3.546 & 3.541 & 3.536\\
$\Phi$ & Hcp & 3.646 & 3.543 & 3.548 & 3.450 & 3.430 & 3.394\\
& Octa & 3.698 & 3.645 & 3.669 & 3.593 & 3.569 & 3.539\\
& Tetra-II & 3.691 & 3.610 & 3.627 & 3.498 & 3.357 &
3.302\\\hline\hline
\end{tabular}
\end{table}

First, we focus our attention to the hydrogen adsorption in the coverage
regime 0%
$<$%
$\Theta\mathtt{\leq}$1.0. Keeping in mind the above-mentioned fact that the
on-surface top and bridge, as well as the subsurface tetra-I site, is unstable
for H adsorption in this coverage regime, our systematic calculations have
been carried out for the remaining four high-symmetry adsorption sites, i.e.,
the on-surface fcc and hcp sites, together with the subsurface tetra-II and
octa sites. The calculated binding energies $E_{b}$ of H on these four
on-surface and subsurface sites, with respect to the free atomic hydrogen, are
illustrated in Fig. 3 and summarized in Table II for different hydrogen
coverage in the regime 0$<$$\Theta\mathtt{\leq}$1.0. One can see that the
binding energy for the on-surface adsorption is always larger than for the
subsurface adsorption. Thus, in the coverage regime 0$<$$\Theta\mathtt{\leq}%
$1.0 the hydrogen on-surface adsorption is more energetically favourable than
the subsurface adsorption, which indicates that the ad-H's will not
spontaneously penetrate into the subsurface in the case of pure on-surface
adsorption. For the on-surface adsorption the fcc site is more stable than the
hcp site, while for the subsurface adsorption the octa site is more favourable
than the tetra-II site. Also, one can see from Fig. 3 that the binding energy
increases with hydrogen coverage for all the four adsorption sites, which
indicates a prominent attraction among the on-surface (or subsurface) ad-H's
and implies a tendency to form H islands or clusters on the Mg(0001) surface
(or subsurface) at 0%
$<$%
$\Theta\mathtt{<}$1.0. On the other hand, considering the binding energy
$E_{b}$ of hydrogen adatom with respect to the half of the binding energy for
the H$_{2}$ molecule (the experimental value of $2.38$ eV), the present
calculations predict that in the whole coverage range we considered in this
subsection, the atomic hydrogen on-surface (as well as subsurface) adsorption
is stable. In addition, interestingly, the binding energy difference between
the on-surface fcc and hcp sites, as well as between the subsurface octa and
tetra-II sites, displays a noticeable increases with hydrogen coverage, which
implies a substrate-induced anisotropy in the hydrogen-metal chemical bonding.
Note that the changes of the binding energy $E_{b}$ from $0.11$ to $0.25$ ML
is not well linear with the coverage and the markable deviation occurs at
$0.33$ ML. This is a result of the effective interaction between the surface adsorbates.

To further clarify our observation that the hydrogen adsorbates tend to form
clusters on the Mg(0001) surface at 0$<$$\Theta\mathtt{<}$1.0, here as one
typical example, we consider two kinds of arrangements for H$_{\text{fcc}}$
adsorbates at the same coverage $\Theta$=0.33. These two adsorbate
arrangements are shown in Fig. 4. The left panel in Fig. 4 corresponds to case
of a $p$($\sqrt{3}\mathtt{\times}\sqrt{3}$) surface cell, while the right
panel gives one selective adsorbate configuration produced from the
$p$(3$\mathtt{\times}$3) surface cell. Obviously, the hydrogen adatoms in the
right panel are more clustered than those in the left panel. Remarkably, the
calculated binding energy is $E_{b}$=$2$.$329$ eV for the left panel and
$E_{b}$=$2$.$380$ eV for the right panel, which clearly shows the preference
for the formation of hydrogen clusters on the Mg(0001) surface. Note that
1$\mathtt{\times}$1 H islands are unfavourable because of the electrostatic
H-H repulsion, which will be discussed in Sec. III C. \begin{figure}[ptb]
\includegraphics[angle=0,width=0.5\textwidth]{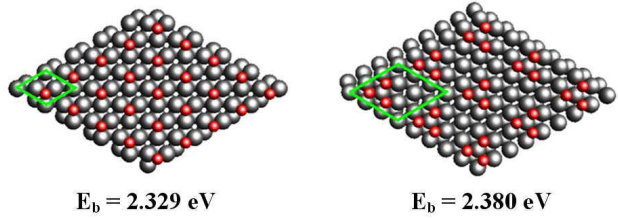} \caption{(Color
online). Two kinds of configurations for H$_{\text{fcc}}$ adsorbates at the
same coverage $\Theta\mathtt{=}0$.$33$. The hydrogen adatoms in the right
panel are arranged to be clustered, while in the left gives an uniform
distribution. The calculated binding energies show that the adsorption in the
right panel is more stable.}%
\label{fig4}%
\end{figure}

Table III \begin{table}[ptb]
\caption{ The calculated adsorbate height h$_{\text{H-Mg}}$, the bond length
($R_{a}$) and the interlayer relaxation ($\Delta_{12}$) for different coverage
of atomic hydrogen adsorption on Mg($0001$) surface.}
\begin{tabular}
[c]{ccccccc}\hline\hline
Coverage & \multicolumn{2}{c}{h$_{\text{H-Mg}}$} & \multicolumn{2}{c}{R$_{a}$
(\AA )} & \multicolumn{2}{c}{$\Delta_{12}(\%)$}\\
$\Theta$ & \ \ Fcc\ \ \  & \ \ Hcp\ \ \  & \ \ \ Fcc \ \  & \ \ \ Hcp \ \  &
\ \ \ Fcc \ \  & \ \ Hcp \ \ \\\hline
0.11 & 0.917 & 0.901 & 2.012 & 2.034 & 2.025 & 2.730\\
0.25 & 0.822 & 0.803 & 2.014 & 2.015 & 1.916 & 2.108\\
0.33 & 0.817 & 0.818 & 1.984 & 1.987 & 1.890 & 1.978\\
0.5 & 0.888 & 0.871 & 2.010 & 2.013 & 1.764 & 2.104\\
0.75 & 0.856 & 0.854 & 2.031 & 2.035 & 1.108 & 1.818\\
1.0 & 0.837 & 0.835 & 2.032 & 2.045 & $-$0.450 & 0.323\\\hline\hline
\end{tabular}
\label{Table3}%
\end{table}presents the calculated results for the relaxed atomic structure,
including the height h$_{\text{H-Mg}}$ of H above the surface, the H--Mg bond
length $R_{a}$, and the topmost interlayer relaxations $\Delta_{12}$ for
various coverage with H in the fcc and hcp sites. One can see that the
adsorption of hydrogen on Mg(0001) induces notable changes in the interlayer
distance of the substrate. In fact, the value of $\Delta_{12}$ monotonically
decreases with hydrogen coverage for both fcc and hcp adsorption. For the fcc
adsorption, in particular, the value of $\Delta_{12}$ even becomes negative at
$\Theta$=1.0, which means that the topmost interlayer relaxation changes from
expansion [about 2.0\% for clean Mg(0001) surface] to contraction. This
reflects the strong influence of the H adsorbates on the neighboring Mg atoms,
and thus results from important redistribution of the electronic structure.
Thus, our results verify that the hydrogen adsorption causes the Mg(0001)
outmost layer separation to relax back to something close to its
\textquotedblleft ideal\textquotedblright\ bulk value. Concerning the H-Mg
bond length $R_{a}$ at different hydrogen coverage, one can see from Table III
that for both fcc and hcp adsorption, the H-Mg bond length varies around 2.0
\r{A}\ very little with increasing $\Theta$. In particular, the calculated
results of $R_{a}$ by using the same $p(2\mathtt{\times}2)$ surface model vary
only within an amplitude of 0.02 \r{A}(0.03 \r{A}) for fcc (hcp) site. The
short bond length $R_{a}$ implies a strong interaction between H and Mg atoms.
Note that the value of $R_{a}$ for H$_{\text{fcc}}$ is slightly shorter than
that for H$_{\text{hcp}}$, which is consistent with the fact that the fcc site
is more stable than the hcp site for on-surface adsorption.

We turn now to analyze the electronic properties of the H/Mg(0001) system by
first considering the work function $\Phi$, which is plotted in Fig. 5 and
summarized in Table II for different hydrogen coverage. For the clean Mg(0001)
our calculated $\Phi$ has a typical value of 3.728 eV, which is well
comparable to the previous calculations. From Fig. 5, it can be seen that the
work function steadily decreases (within a relatively small variation range)
with H coverage for both on-surface and subsurface adsorption. On the whole,
the variation amplitude of the work function is relatively small, which can be
associated with the small adsorption distance for the H species. The
decreasing linetype and amplitude of $\Phi$ shown in Fig. 5 as a function of
$\Theta$ depend on the adsorption configuration. For the on-surface
adsorption, one can see that the work function for the hcp adsorption is
always lower than that for the fcc adsorption, and their difference increases
with $\Theta$, which means that the surface charge polarization effect is more
prominent for the hcp adsorption than for the fcc adsorption in the whole
coverage considered. For the subsurface adsorption, In a similar manner, the
octa-adsorbed work function is always lower than that for the tetra-II
adsorption, and their difference also increases with $\Theta$. Interestingly,
with increasing the hydrogen coverage, the value of work function for the
on-surface fcc adsorption approaches to have the same value as that for the
subsurface octa adsorption. \begin{figure}[ptb]
\includegraphics[angle=0,width=0.4\textwidth]{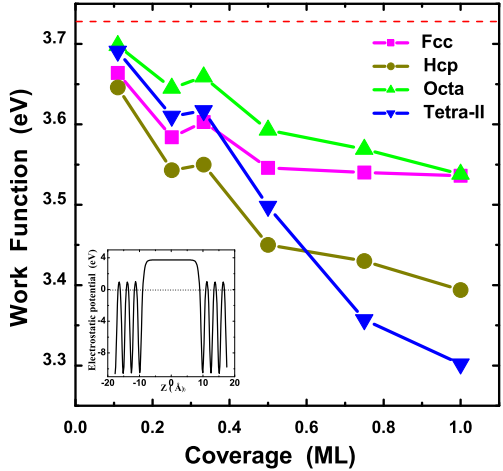} \caption{(Color
online). The calculated work function $\Phi$ versus the coverage for the H
atom adsorption in different sites. The dashed line shows the value of $\Phi$
for clean Mg(0001) substrate. As an example, the inset shows the
planar-averaged electrostatic potential of clean Mg(0001) slab, with the Fermi
energy set at zero.}%
\label{fig5}%
\end{figure}

To gain more insight into the nature of Mg-H bonding during the hydrogen
adsorption onto Mg(0001) surface, we now analyze our results by means of the
electron density difference $\Delta n$($\mathbf{r}$), which is obtained by
subtracting the electron densities of noninteracting component systems,
$n_{\text{Mg(0001)}}$($\mathbf{r}$) +$n_{\text{H}}$($\mathbf{r}$), from the
density $n$($\mathbf{r}$) of the H/Mg(0001) system, while retaining the atomic
positions of the component systems at the same location as in H/Mg(0001).
Figure 6(a) and 6(b) \begin{figure}[ptb]
\includegraphics[angle=0,width=0.5\textwidth]{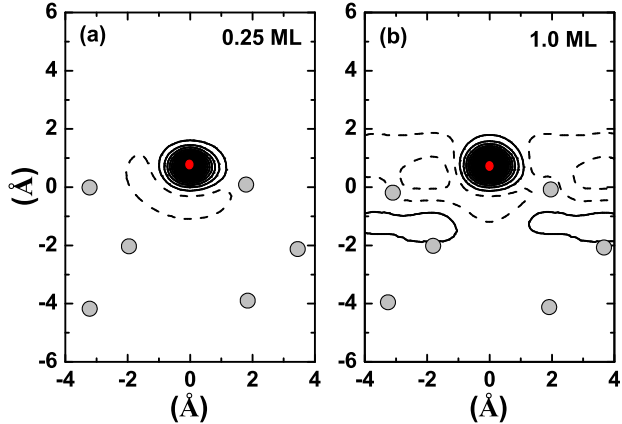} \caption{(Color
online). Contour plot of the charge density difference $\Delta n(\mathbf{r})$
for the on-surface H/Mg(0001) slab with hydrogen coverage (a) $\Theta$=0.25
and (b) $\Theta$=1.$0$. The solid (dashed) lines stand for the electron
accumulation (depletion) regions. The red (gray) points stand for H and Mg
atoms. The contour spacing is 0.02 electrons/\AA $^{3}$.}%
\label{fig6}%
\end{figure}present the contour plots of $\Delta n$($\mathbf{r}$) for $\Theta
$=0.25 and 1.0, respectively. One can see that the charge redistribution
mainly occurs at the surface and involves in the H adatom and the topmost Mg
atoms. It is apparent that upon adsorption, electrons flow from Mg $sp$
metallic state into H $1s$ state, resulting in a depletion of the surface
metallic electrons. With the increase in hydrogen coverage, it shows that (i)
more Mg $sp$ electrons transfer to the localized H $1s$ orbital, implying that
ionicity of the Mg-H bonding increases with hydrogen coverage, and (ii) not
only the topmost but also the second Mg atomic layer is prominently influenced
by increasing the hydrogen coverage, which can be clearly seen by the charge
redistribution shown in Fig. 6(b). On the other hand, there is also a weak but
important covalent component in Mg-H chemical bonding (see Fig. 7 below for
details). The unique signiture of this covalency occurred in Fig. 6 is the
orientation of the charge accumulation (around H adatom), which, due to the
hybridization of surface Mg $sp$ and H $1s$ states, tends to point along the
Mg-H bond. Thus, one can see that the chemical bonding between the surface Mg
atom and H adatom is a mixture of ionic and covalent bonding. For pure
MgH$_{2}$ with rutile structure, recent experimental and theoretical studies
have shown that the bulk Mg-H chemical bonding also has a mixed ionic/covalent
nature. Another fact shown in Fig. 6 is that like the other hydrogen/metal
systems, the influence of the adsorbed Mg(0001) surface is rapidly screened
out on going into the bulk. The bonding character of the inner Mg layers (from
the second layer for $\Theta$=0.25 and from the third layer for $\Theta$=1.0)
is essentially identical to the bulk case, which is typically metallic with a
fairly constant charge density between the atoms with slight directional
bonding along the body diagonals. This can be clearly seen in Fig. 6 which
shows negligibly small changes in the interior of the nine-layer Mg slab at
both low ($\Theta$=0.25) and high ($\Theta$=1.0) coverage.

Figures 7(a) and 7(b) \begin{figure}[ptb]
\includegraphics[angle=0,width=0.5\textwidth]{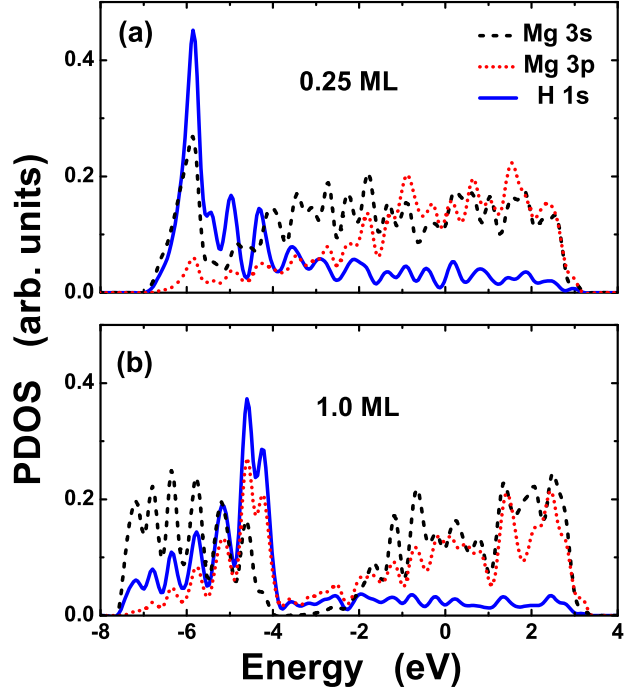} \caption{(Color
online). The PDOS for the on-surface H$_{\text{fcc}}$ atoms and the topmost Mg
layer at (a) $\Theta$=0.25 and (b)$\ \Theta$=1.0. The Fermi level is set at
zero.}%
\label{figure7}%
\end{figure}show the orbital-resolved PDOS for the on-surface H$_{\text{fcc}}$
layer and the topmost Mg layer at $\Theta$=0.25 and $\Theta$=1.0,
respectively. The Fermi energy has been set at zero. At low coverage ($\Theta
$=0.25), the narrow peak around $-$6.0 eV with large amplitude denotes H $1s$
state, which, as shown in Fig. 7(a), mainly hybridizes with the $3s$ state of
the outmost Mg atoms. Whereas, the hybridization between H $1s$ and Mg $3p$
states is negligibly small. By comparing the low-coverage adsorbed [Fig. 7(a)]
and clean [Fig. 2(b)] Mg(0001) surfaces, one can see that the main change for
the surface Mg layer upon the low-coverage hydrogen adsorption is the distinct
enhancement of its $3s$ PDOS at the Mg valence band edge (around $-$6.0 eV)
due to the bonding with H $1s$ state. At this moment, it is interesting to
question why the electronic bonding of the hydrogen adatom and the surface Mg
atoms occurs at the Mg valence band edge in the case of low adsorption
coverage. The reason is simply because this band edge, as shown in Fig. 2(b),
is dominated by the Mg $3s$ state, while the other part of the Mg valence band
is a mixture of Mg $3s$ and $3p$ states. At low coverage, the H adatom has the
low coordinates, which makes it favourable for the H($1s$)-Mg($3s$) chemical
bonding. Thus, at low coverage the H adatom chooses to bond at the Mg valence
band edge as shown in Fig. 7(a). With increasing hydrogen coverage [Fig. 7(b)
for $\Theta$=1.0], three prominent changes involving the H-Mg chemical bonding
occur: (i) the peak in the H $1s$ PDOS is broadened and shifts upward to go
into the interior of the Mg valence band. These changes for the H $1s$ PDOS is
due to the fact that at as high coverage as $\Theta$=1.0, the H adatom is
highly coordinated, which drives the H $1s$ state to bond with not only the
$3s$ state but also the $3p$ states of the topmost Mg atoms. Since the Mg $3p$
states lies mainly in the interior of the valence band, thus the H $1s$ state
has to shift up in energy to overlap with the Mg $3p$ states; (ii) Compared to
case of $\Theta$=0.25, the hybridization of H $1s$ and Mg $3p$ states is
distinctly enhanced in the case of $\Theta$=1.0. In particular, the main peak
around $E$=$-$4.5 eV in the H $1s$ PDOS in Fig. 7(b) is a result of the
hybridization between H $1s$ and Mg $3p$ states. The reason for this
increasing H($1s$)-Mg($3p$) hybridization at high coverage has been clarified
in (i); (iii) The distribution of Mg $3s$ and $3p$ states in the energy region
$-$4.0%
$<$%
$E$%
$<$%
$-$2.0 eV tends to vanish due to a large weight transfer of these states to
more lower energy, which is caused by the formation of the bonding and
antibonding states between H $1s$ and Mg $sp$ atomic orbitals. This is most
prominent for Mg $3s$ state. In fact, one can see from Fig. 7(b) that in the
energy interval $-$7.0%
$<$%
$E$%
$<$%
$-$6.0 eV there is a large filling of the surface Mg $3s$ state, which is
empty in this energy region in the cases of clean [Fig. 2(b)] and low-coverage
[Fig. 7(a)] Mg(0001) surfaces. Obviously, this state-transfer toward more
lower energy in increasing adsorption coverage will gain the energy, which
overcompensates the energy that is costed by the elevation of the H $1s$ state
when increasing adsorption coverage. The net result is that the adsorption of
the on-surface H$_{\text{fcc}}$ at $\Theta$=1.0 is more stable than at
$\Theta$=0.25, as has been shown in Fig. 3.

After clarifying the chemical bonding properties for the most stable hydrogen
adsorption configuration, i.e., the on-surface fcc site, we turn now to study
the diffusion energetics of the H adatom when it undergoes the in-plane
diffusion and the inter-plane penetration (from surface to subsurface).
Diffusion of atomic hydrogen after on-surface dissociation of H$_{2}$ is an
elementary process during the surface hydrogenation process. Also, the
hydrogen diffusion plays an important role in understanding many catalytic
reaction. Here, by using the DFT total energy calculation, we report our
numerical results of the energy barriers for atomic H diffusion and
penetration in the H/Mg(0001) system.

Using the nudged elastic band (NEB) method, which will find the saddle points
and minimum energy paths on complicated potential surface, we have calculated
the surface diffusion-path energetics of atomic hydrogen. The results at
$\Theta$=0.25 and $\Theta$=1.0 are shown in Fig. 8(a). Note that during
calculation the number of atoms keeps invariant. Thus it also reveals in Fig.
8(a) the relative stability of H adsorption among various on-surface sites and
the corresponding hydrogen binding energy differences. Our calculated
diffusion barrier from fcc to hcp site is 0.33 eV at $\Theta$=0.25 and 0.18 eV
at $\Theta$=1.0. The hcp site is less stable than the fcc site within the
coverage 0%
$<$%
$\Theta\mathtt{\leq}$1.0. Thus the on-surface diffusion barrier from hcp to
fcc site presents an activation barrier with the value of 0.30 eV at $\Theta
$=0.25 and 0.08 eV at $\Theta$=1.0. Therefore, we find that both the diffusion
barrier from fcc to hcp site and activation diffusion barrier from hcp to fcc
site is greatly decreased with coverage, which implies that at the coverage
close to $\Theta$=1.0, the H/Mg(0001) surface at low temperature may display a
mixed (disordered) phase that while the fcc sites are mainly occupied, there
exist a small amount of ad-H's residing on the hcp sites due to thermal
diffusion. On the other hand, the penetration-path energetics of surface
ad-H's to the subsurface sites also sensitively depends on the coverage. Since
the most favourable on-surface and subsurface adsorption sites are the fcc and
octa sites respectively, also since there is no geometric obstacle between the
neighboring fcc and octa sites, it is natural to speculate that at low
temperature, the ad-H's on the surface fcc sites may penetrate to the
neighboring subsurface octa sites through overcoming a relatively low barrier
in the coverage range 0%
$<$%
$\Theta\mathtt{\leq}$1.0. Thus, here we consider the energetics of this H
penetration path. For this purpose, we fully relax the topmost three Mg layers
step by step to search for the transition state with high energy. The
calculated penetration path and the energy barrier from the on-surface fcc to
subsurface octa site at two coverage ($\Theta$=0.25 and $\Theta$=1.0) is shown
in Fig. 8(b), in which the horizontal coordinate indicates the penetration
distance of the H atom with respect to the topmost Mg layer. At the low
coverage of $\Theta$=0.25, the calculated barrier for fcc$\rightarrow$octa
penetration is as large as 0.45 eV. The transition state, i.e., the atomic
geometry of the energy maximum in the penetration path, correspond to the
hydrogen atom in the surface Mg layer. When the coverage is increased to
$\Theta$=1.0, as shown in Fig. 8(b), it is surprisingly found that the barrier
is further increased to 0.75 eV. This is unlike the case of surface diffusion
shown in Fig. 8(a), in which the energy barrier is weakened by increasing the
adsorption coverage. Therefore, in contrast with one's intuitive expectation,
our calculated results in Fig. 8(b) shows that the probability of
low-temperature interlayer penetration is negligibly small in the coverage
range 0%
$<$%
$\Theta\mathtt{\leq}$1.0. \begin{figure}[ptb]
\includegraphics*[angle=0,width=0.5\textwidth]{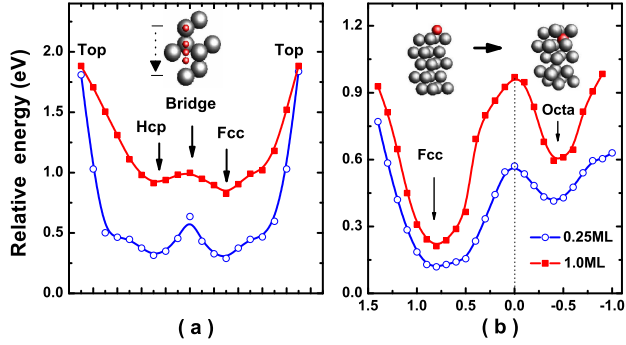}\newline%
\caption{(Color online). (a) The surface diffusion and (b) penetration (from
on-surface fcc to subsurface octa site) energetics for the H adatom on the
Mg(0001) with the coverage of 0.25 and 1.0 ML, respectively. The insets
schematically plot atomic geometries along the diffusion and penetration
pathes. }%
\label{FIG:figure8}%
\end{figure}

\subsection{Atomic hydrogen adsorption at 1.0$\mathtt{<}\Theta\mathtt{\leq}%
$2.0}

From above discussions, one can see that in the coverage range 0%
$<$%
$\Theta\mathtt{\leq}$1.0, the most stable adsorption site for H/Mg(0001)
system is the on-surface fcc site. To penetrate into the interior of the
magnesium substrate, the hydrogen adatom has to overcome a large-amplitude
($\sim$0.75 eV) energy barrier, which has been shown in Fig. 8(b) for the
energy path from the on-surface fcc to the subsurface octa site. This means
that at low temperature, the H adatoms will mostly reside on the Mg(0001)
surface and a spontaneous penetration process cannot occur at 0%
$<$%
$\Theta\mathtt{\leq}$1.0. Then, one may naturally ask an important question:
what will happen when the hydrogen coverage is more than 1.0 ML, or what is
the saturation coverage for surface adsorption? This question is closely
related to the clear understanding of the surface hydrogenation of magnesium
and is answered in the following discussions by calculating the hydrogen
adsorption energetics in the coverage range 1.0$\mathtt{<}\Theta\mathtt{\leq}%
$2.0. One should notice that up to now the exact knowledge of saturation
coverage is still unreasonable experimentally because of the limitation of the
resolution and device factors. Theoretical study and prediction of surface
hydrogenation are thus still particularly enlighting at the present stage.

To study the hydrogen adsorption in the coverage range 1.0$\mathtt{<}%
\Theta\mathtt{\leq}$2.0 with allowable computation time, we choose a
$p$($2\times2$) supercell to construct four coverage configurations, i.e.,
$\Theta$=1.25, 1.5, 1.75, and 2.0. The initial positions of H atoms for
relaxation calculation are set such that one ML of H atoms are placed on the
surface fcc sites of Mg(0001), then the remaining ($\Theta\boldsymbol{-}$1.0)
ML of H's are placed on the surface hcp sites of Mg(0001), see Fig. 9(a)
\begin{figure}[ptb]
\includegraphics[angle=0,width=0.5\textwidth]{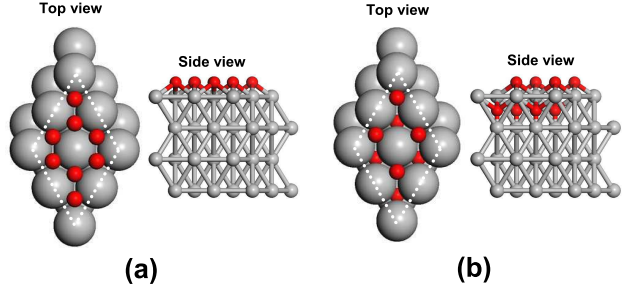} \caption{(Color
online). (a) The initial struture (4fcc+4Hcp) and (b) the end structure
(4fcc+4Tetra-I ) in the relaxation of H/(2$\times$2)-Mg(0001) system with
2.0ML coverage.}%
\label{figure9}%
\end{figure}for the top and side view of one selective initial atomic geometry
($\Theta$=2.0). This choice of initial geometries is simply motivated by the
above low-coverage result that the on-surface fcc and hcp sites are more
stable than the subsurface tetra-II and octa sites. Then, the total-energy
relaxation calculation is carried out in the same manner as that in the above
discussion. After atomic relaxation, prominently, we find that the H atoms
initially placed on the surface hcp sites spontaneously move to the subsurface
tetra-I site [see Fig. 9(b)] without encountering any penetration barrier. The
driving force for this spontaneous penetration process may be due to the
mutual repulsion of high-coverage H adatoms. This reasoning is consistent with
the partially-ionic chemical bonding between H and Mg atoms, which results in
a negative charge carried by H's and thus generates a repulsion among these
adsorbates. Thus, we arrive at a conclusion that within the coverage range
1.0$\mathtt{<}\Theta\mathtt{\leq}$2.0, the surface and subsurface of Mg(0001)
are coadsorbed, with 1.0 ML of H's residing on the surface sites and
($\Theta\mathbb{-}$1.0) ML of H's occupying the subsurface tetra-I sites.

\begin{table}[ptb]
\caption{ Calculated binding energies E$_{b}$\ (in eV) and the
interlayer relaxation ($\Delta_{12}$ and $\Delta_{23}$) for atomic
hydrogen adsorption on Mg(0001) surface at coverage of 1.0 $<$
$\Theta\leq$2.0.}
\begin{tabular}
[c]{ccccc}\hline\hline
$\Theta$ & \ 1.25 \ \  & \ \ \ 1.5 \ \ \  & \ \ 1.75 \ \ \  & \ \ \ 2.0 \ \\
$\Delta_{12}$(\%) & 2.266 & 5.738 & 9.373 & 13.30\\
$\Delta_{23}$(\%) & 0.179 & 1.452 & 2.628 & 3.50\\
$E_{b}$ (eV) & 2.504 & 2.518 & 2.515 & 2.511\\\hline\hline
\end{tabular}
\end{table}Even more interestingly, we find that this coadsorption of both
on-surface and subsurface sites is energetically more favourable than the pure
on-surface or subsurface adsorption. This can be seen from Table IV, in which
the calculated hydrogen binding energies for $\Theta$=1.25, 1.5, 1.75 and 2.0
(with the above-mentioned initial geometries) are listed. By comparing with
the binding energy data listed in Table II, one can see from Table IV that the
hydrogen high-coverage coadsorption at 1.0%
$<$%
$\Theta\mathtt{\leq}$2.0 [with 1.0 ML H's residing on the surface sites and
($\Theta-$1.0) ML H's occupying the subsurface tetra-I sites] is more stable
than the pure on-surface or subsurface adsorption at 0%
$<$%
$\Theta\mathtt{\leq}$1.0. Note that there also exists many stable coadsorption
configurations in the coverage range 0%
$<$%
$\Theta\mathtt{\leq}$1.0. We have systematically calculated the total energies
of these low-coverage coadsorption configurations. For example, for
$\Theta\mathtt{=}$0.5 we have used the $p$($2\times2$) supercell to construct
the following three coadsorption geometries: (i) one H is on the surface fcc
site and the other one H is on the subsurface tetra-I site; (ii) one H is on
the surface fcc site and the other one H is on the subsurface tetra-II site;
(iii) one H is on the surface fcc site and the other one H is on the
subsurface octa site. For the other coverage satisfying 0%
$<$%
$\Theta\mathtt{\leq}$1.0, the coadsorption configurations are constructed in a
similar manner. Our systematic calculations clearly show that within the
coverage range 0%
$<$%
$\Theta\mathtt{\leq}$1.0, the coadsorption is less stable than the pure
on-surface fcc adsorption. Combining the results for both 0%
$<$%
$\Theta\mathtt{\leq}$1.0 and 1.0%
$<$%
$\Theta\mathtt{\leq}$2.0, therefore, we get a critical coverage value of
$\Theta\mathtt{=}$1.0 for hydrogen adsorption on Mg(0001). Below this value
the ad-H's occupy the on-surface fcc sites, while above this value (1.0%
$<$%
$\Theta\mathtt{\leq}$2.0), 1.0 ML of H's occupy the on-surface fcc sites and
the remaining ($\Theta-$1.0) ML of H's occupy the subsurface tetra-I sites.
Here, the occupation of the subsurface tetra-I sites can be understood as a
result of the spontaneous penetration process of the on-surface H$_{\text{hcp}%
}$ adatoms, provided the on-surface fcc sites are fully occupied. As has been
mentioned above, the driving force for this spontaneous penetration process is
possibly from the repulsion of the neighboring on-surface ad-H dipoles.
Dipole-dipole repulsion is a mechanism that has been proposed to explain how
surface oxidation begins. Remarkably, our present results clearly show a
similar mechanism responsible to the incipient hydrogenation of Mg(0001) surface.

Now let us turn back to Table IV, in which, besides the H adsorption energy,
also presents the first and second interlayer relaxations ($\Delta_{12}$ and
$\Delta_{23}$), as well as the H-Mg bond length $R_{a}$, for the relaxed
atomic structures of four coverage in the region 1.0%
$<$%
$\Theta\mathtt{\leq}$2.0. Prominently, one can see that the coadsorption of
hydrogen on Mg(0001) induces distinct changes in the interlayer distance of
the substrate. In particular, the topmost interlayer relaxation changes from
contraction $\Delta_{12}$=$-$0.45$\%$ at $\Theta\mathtt{=}$1.0 to expansion
$\Delta_{12}$=13.3$\%$ at $\Theta\mathtt{=}$2.0. This large change of atomic
relaxations upon H coadsorption is largely ascribed to the occupation of
ad-H's at the subsurface tetra-I sites, which together with the surface
ad-H's, form a H-Mg-H sandwich structure and thus breaks metallic bonding
between the first and second Mg(0001) atomic layers. Because of this large
interlayer expansion one can speculate that this H-Mg-H sandwich structure is
a precursor in the process of H-induced embrittlement and ablation of
Mg(0001). Needless to say, larger interlayer expansion costs more energy,
which will counteract the formation of H/Mg(0001) structures of more higher
coverage. This statement is supported by the result shown in Table IV that the
binding energy $E_{b}$ arrives at its maximum at $\Theta\mathtt{=}$1.5. Based
on this observation, in this paper we do not consider the coverage region of
$\Theta\mathtt{>}$2.0.

To help understand the hydrogenation of Mg(0001), which is characterized by
the occurrence of surface hydride, it is instructive to compare the electronic
properties of the present H-Mg-H sandwich structures and the bulk hydride
MgH$_{2}$. For this we have calculated the ground-state properties of the bulk
MgH$_{2}$ with rutile structure. Our optimized lattice parameters and internal
coordinate parameters are in good agreement with the experimental data
\cite{Noritake2002}. Then, we calculate and plot in Fig. 10(a)
\begin{figure}[ptb]
\includegraphics[angle=0,width=0.5\textwidth]{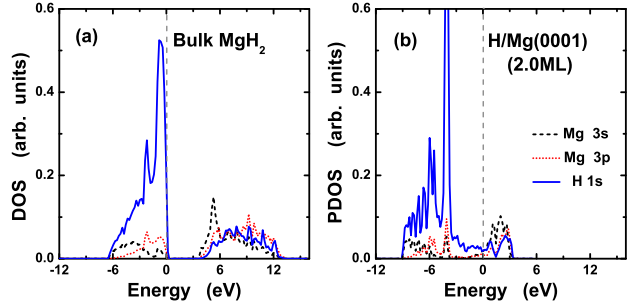} \caption{(Color
online). (a) The DOS of bulk MgH$_{2}$ and (b) the projected DOS for H-Mg-H
sandwich structure of H/Mg(0001) at coverage $\Theta\mathtt{=}$2.0. The energy
is measured with reference to the Fermi level.}%
\label{fig10}%
\end{figure}the orbital-resolved PDOS of H and Mg in MgH$_{2}$. One can see
that MgH$_{2}$ is an insulator with a band gap of 4.0 eV. The valence band is
dominated by H $1s$ and the conduction band mainly consists of Mg $sp$ states,
which means that the Mg-H bonds in MgH$_{2}$ are mainly ionic. On the other
hand, Fig. 10(a) also shows that a small amount of Mg $sp$ states in the
valence band hybridize with the H $1s$ state and contribute some covalent
character to the bonding. It has been experimentally measured
\cite{Noritake2002} that the ionic charge of Mg and H atoms in MgH$_{2}$ are
represented Mg$^{1.91+}$ and H$^{0.26-}$, which implies that the Mg and H
atoms in MgH$_{2}$ are ionized in-equivalently. In other words, Mg is ionized
almost as Mg$^{2+}$, while H is very weakly ionized. Figure 10(b) plots the
orbital-resolved PDOS for the present H-Mg-H sandwich structure of H/Mg(0001)
at coverage $\Theta\mathtt{=}$2.0. In plotting Fig. 10(b) we have used the
same number of H and Mg atoms as in Fig. 10(a). It can be seen that although
there is an understandable difference in the spectral positions of H $1s$ and
Mg $sp$ states (relative to the Fermi energy) between the H-Mg-H sandwich
structure of H/Mg(0001) and the bulk MgH$_{2}$, the PDOS peak features for the
two cases are yet very similar. For example, there are two prominent H $1s$
peaks below $E_{F}$ for both the H-Mg-H sandwich structure of H/Mg(0001) and
the bulk MgH$_{2}$; the energy distance between these two peaks is also very
comparable for the two cases. The band charge distribution around these two H
$1s$ peaks (not shown here) also reveals the similarity between the two cases.
Therefore, we expect that it is possible to form on Mg(0001) the surface
hydride with the similar chemical stoichiometry as the bulk MgH$_{2}$. From
this perspective we can see that Mg(0001) is a good case for studying the
hydrogenation (and dehydrogenation) performance.

\section{Conclusion}

In summary, we have systematically investigated the adsorption of atomic
hydrogen on Mg(0001) surface and subsurface, as well as the energy barriers
for atomic H diffusion and penetration in these systems through
first-principles DFT-GGA calculations. We have considered a wide range of
coverage from 0.11 to 2.0 ML by using different surface models [i.e.,
$p$(3$\times$3), p(2$\times$2), $p$($\sqrt{3}\mathtt{\times}\sqrt{3}$) and
$p$(1$\times$1) surface unit cells] for adsorption in the on-surface fcc and
hcp sites, as well as in the subsurface tetra-I, tetra-II, and octa sites. In
the coverage range 0%
$<$%
$\Theta\mathtt{\leq}$1.0, the most stable among all possible pure adsorbed
sites, as well as coadsorbed sites, is the on-surface fcc site, followed by
the hcp site. The atomic geometry, the work-function change, the charge
density distribution, and the electronic structure upon the H adsorption have
also been studied, which consistently show the fundamental influence by the
ionic as well as covalent bonding between the H\ adatom and surface Mg atoms.
Remarkably, this influence in the energetics is increased with increasing the
H coverage, which is highly interesting. For instance, the increase in the H
binding energy for the fcc or hcp site with $\Theta$ in the low coverage range
(0%
$<$%
$\Theta\mathtt{<}$1.0) implies the effective attraction between the H
adsorbates, which will make it favourable for the formation of the hydrogen
island or cluster \textit{at low coverage less than one monolayer}. It should
be stressed that the 1$\mathtt{\times}$1 H island is unfavourable because of
the electrostatic H-H repulsion. Furthermore, in the low coverage region 0%
$<$%
$\Theta\mathtt{\leq}$1.0 we have also calculated the surface diffusion as well
as surface-to-subsurface penetration path energetics. For the on-surface
diffusion, it has been found that the fcc and hcp sites are the two local
minima, while the bridge is saddle point along the diffusion path. The
activation barrier for the surface diffusion from hcp to fcc site is 0.30 eV
at $\Theta\mathtt{=}$0.25 and 0.08 eV at $\Theta\mathtt{=}$1.0, implying that
with increasing the hydrogen coverage in the region 0%
$<$%
$\Theta\mathtt{\leq}$1.0, the surface diffusion tends to be more easier. On
the contrary, the activation barrier for the penetration from the on-surface
fcc to the subsurface octa site is largely increased with increasing the
hydrogen coverage. Thus, we have found that although the octa site is most
stable for the pure subsurface adsorption, it is actually very hard to be
reached at for H adatoms after dissociation of H$_{2}$ on Mg(0001) surface.
This fact, together with the observation that the binding energy for the most
stable on-surface fcc site does not saturate at 0%
$<$%
$\Theta\mathtt{\leq}$1.0, has motivated us to study hydrogen adsorption
properties at more higher coverage of 1.0%
$<$%
$\Theta\mathtt{\leq}$2.0.

In the coverage range 1.0%
$<$%
$\Theta\mathtt{\leq}$2.0 and starting with the initial adsorption
configurations that 1.0 ML of H's are placed on the surface fcc sites and
($\Theta-$1.0) ML of H's are placed on the surface hcp sites, we have found
that after atomic relaxation, while the fcc H's keep the same sites, the hcp
H's however undergo spontaneous penetration to the subsurface tetra-I sites,
which are otherwise the unstable sites for the pure subsurface adsorption.
This spontaneous penetration phenomenon is ascribed to the strong
electrostatic H-H repulsion at 1.0%
$<$%
$\Theta\mathtt{\leq}$2.0. More interestingly, this final coadsorption
configuration with 1.0 ML of H's residing on the surface fcc sites and the
remaining ($\Theta\mathbb{-}$1.0) ML of H's occupying the subsurface tetra-I
sites is believed to be the most stable coadsorption configuration, after
having compared its formation energy to those of a large number of other
coadsorption configurations. In addition, the resultant H-Mg-H sandwich
structure for this most stable coadsorption configuration has been shown to
display similar DOS spectral features to the bulk hydride MgH$_{2}$. Thus, we
believe the present calculated results may greatly help understand the
incipient hydrogenation of Mg(0001) surface.\ \ \

\begin{acknowledgments}
P.Z. was supported by the NSFC under Grants No. 10604010 and No. 60776063.
Y.W. was supported by the NSFC under Grants No. 50471070 and No. 50644041.
\end{acknowledgments}

\end{document}